\documentclass[aps,prl,twocolumn,showpacs]{revtex4}
\usepackage{graphicx,amsfonts,amsmath,amssymb}

\input epsf
\def\b{\begin{equation}}
\def\e{\end{equation}}

\begin{document}
\title{Exploring the phase space of the quantum delta kicked accelerator}
\author{G. Behinaein, V. Ramareddy, P. Ahmadi, and G.S. Summy}
\affiliation{Department of Physics, Oklahoma State University,
Stillwater, Oklahoma 74078-3072}

\begin{abstract}
We experimentally explore the underlying pseudo-classical phase
space structure of the quantum delta kicked accelerator. This was
achieved by exposing a Bose-Einstein condensate to the spatially
corrugated potential created by pulses of an off-resonant standing
light wave. For the first time quantum accelerator modes were
realized in such a system. By utilizing the narrow momentum
distribution of the condensate we were able to observe the
discrete momentum state structure of a quantum accelerator mode
and also to directly measure the size of the structures in the
phase space.
\end{abstract}


\maketitle For more than a century the study of chaotic phenomena
has been recognized as being crucial to developing a fuller
understanding of nature. One aspect of this study which was
missing until relatively recently was experimental scrutiny of
quantum systems which in the classical limit exhibit chaotic
behavior. Theoretical work on this subject had largely
concentrated on the investigation of idealized systems such as the
quantum delta-kicked rotor (QDKR) which were already well known
from extensive work in the classical regime \cite{Lichtenberg}.
The experimental study of this system gained new impetus through
its realization using laser cooled atoms exposed to a corrugated
potential from a pulsed off-resonant standing light wave
\cite{Moore}. This system has subsequently led to many discoveries
in the field of quantum chaos including observation of quantum
resonances \cite{Moore, Ryu}, dynamical localization
\cite{Moor2,Ringot} and quantum diffusion \cite{Ammann,Duffy2004}.
In the theoretical description of the QDKR the effective value of
Planck's constant scales with the time between the pulses
\cite{Oskay2000, Sadgrove}. Therefore, to achieve classical
correspondence in which $\hbar \rightarrow 0$, the time between
pulses needs to be close to zero. However, for technical reasons
this value can not be made arbitrarily small in experiments.

Recently, Fishman, Guarneri, and Rebuzzini (FGR) \cite{Fishman}
have shown that this difficulty can be circumvented for kicking
periods close to a quantum resonance time. These resonance times
\cite{Oskay, darcy3} occur at integer multiples of the Talbot
time, a time interval during which plane waves with certain
equally spaced momenta in the kicking direction can acquire a
phase factor which is an integer multiple of $2\pi$. This is
analogous to the Talbot effect in optics \cite{berry}. Using this
method, FGR have studied the quantum delta-kicked accelerator
(QDKA) which can be created by taking the QDKR and adding a linear
potential along the direction of the standing wave. They showed
that the effective value of Planck's constant scales with the
deviation of the pulse separation time from certain non-zero
resonance times, thus making it feasible to make the effective
Planck's constant very small. Therefore, if the time between
pulses is chosen close to these resonance times, a
pseudo-classical approach can be adopted to study the system.
\begin{figure}[h]
\begin{center} \mbox{\epsfxsize 3.0in\epsfbox{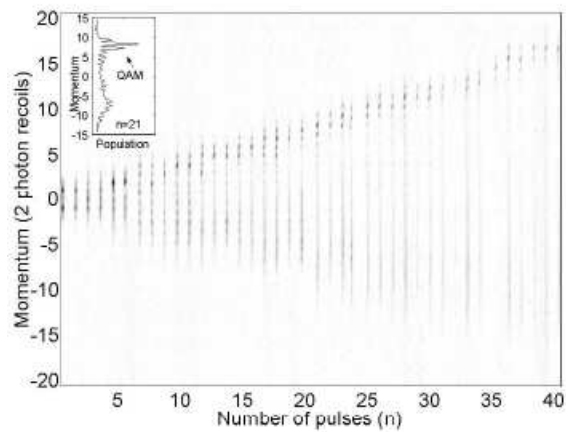}} \end{center}
\caption {Experimental momentum distributions showing a QAM in a
BEC exposed to a series of kicks from a standing light wave in the
free falling frame. The kicking period was 72.2 $\mu$s and the
momentum distributions are displaced as a function of the number
of kicks. The accelerator mode is the collection of momentum
states that appear to move towards the upper right. The inset
shows the momentum distribution for a pulse number of $n$=21.}
\label{momenutm_pulsenumebr}
\end{figure}

Perhaps the simplest way of experimentally realizing the QDKA is
by applying the pulsed standing wave in the direction of gravity
\cite{Oberthaler,Godun}. This experiment has led to the discovery
of quantum accelerator modes (QAMs) near the resonance times. One
of the most important characteristics of QAMs is that they are
comprised of atoms which show a linear momentum growth with pulse
number in a freely falling frame \cite{Oberthaler}. It has been
shown that QAMs are quantum nondissipative counterparts of mode
locking \cite{Buchleitner}. They have also been suggested for use
in the preparation of well defined initial conditions for quantum
chaos experiments \cite{d'Arcy1}. FGR attributed the QAMs to the
existence of stability islands in the pseudo-classical phase
space. These studies have shown that this underlying phase space
has a complex structure which is highly sensitive to the
experimental parameters. However, the broad momentum distribution
of the laser cooled atoms which have been used so far to study
this system have prevented the examination of the local structures
in the phase space.

In this letter we report on the realization of QAMs using a
Bose-Einstein condensate (BEC) of rubidium 87 atoms and the
exploration of the pseudo-classical phase space structure of the
QDKA. Figure\thinspace \ref{momenutm_pulsenumebr} shows
experimentally observed momentum distributions as a function of
the number of standing wave pulses applied to a BEC. This figure
demonstrates that the QAM gains momentum linearly as the number of
pulses increases. Note that this is the first time that it has
been possible to determine that the QAM is made up of several
distinct momentum states as originally postulated in
Ref.\thinspace \cite{Godun}. This quantization of momentum is
observable because the initial momentum uncertainty of the
condensate was much smaller than two photon recoils.

The Hamiltonian of the delta kicked accelerator for an atom with
mass $m$ can be written as, $H = p^2/2m
 + m g' x + {U_{{\rm max}}\over 2}[1+\cos (G x)] \sum_{n} \delta (t -
 n T)$, where $p$ is the atomic momentum along the standing wave,
$g'$ is the component of the gravitational acceleration in the
direction of the standing wave, $G=2k$ the grating vector where
$k$ is the light wave vector, $n$ the pulse number, $T$ the pulse
period, and $U_{{\rm max}}$ is the well depth of the standing wave
due to the light shift. The net effect of the time dependent
potential is to distribute the condensate into different momentum
states via a diffractive process. The population of these momentum
states is determined by $|J_{n'}(\phi_d)|^2$, where
$J_{n'}(\phi_d)$ is an $n'$th order Bessel function of the first
kind with argument $\phi_d = U_{{\rm max}}\triangle t/(2 \hbar)$,
the phase modulation depth \cite{Godun}. It is useful to consider
this system for pulse periods close to integer multiples of the
half Talbot time, $T_{\ell} = \ell \times 2\pi m / \hbar G^2 (=
\ell \times 33.3 \mu s$ for Rb87 atoms). With this restriction the
system can be described by the classical map \cite{Fishman},
\begin{eqnarray}
\theta_{n+1} &=& \theta_n + {\rm {sgn}} (\epsilon) J_n
\nonumber \\
J_{n+1} &=& J_n - K \sin (\theta_{n+1}) + {\rm{sgn}}(\epsilon)
\tau \eta,\label{map}
\end{eqnarray}
where $\epsilon = 2 \pi \ell(T/T_{\ell} - 1)$ is a small number,
$\ell$ is any positive integer number, and $K = |\epsilon|
\phi_d$. The dimensionless $J$ and $\theta$ parameters are defined
as,
\begin{eqnarray}
\theta &=& G~ x ~{\rm {mod}} (2 \pi) \nonumber \\
J_n &=& I_n + {\rm{sgn}}(\epsilon) [\pi \ell + \tau (\beta + n
\eta + \eta /2)] \label{dynamic}
\end{eqnarray}
where $p/(\hbar G) = I /|\epsilon| + \beta$, $\beta$ is the
fractional momentum, $\tau = \hbar T G^2 /m$, and $\eta = m g'
T/(\hbar G)$.
\begin{figure}[h]
\begin{center} \mbox{\epsfxsize 3.0in\epsfbox{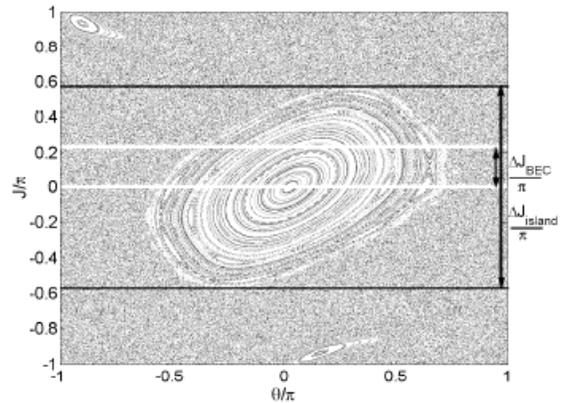}} \end{center}
\caption {Phase space unit cell for the QDKA map of Eq.\thinspace
(\ref{map}) with $\phi_d = 1.4$ and $T = 29.5$ $\mu$s. A stable
fixed point with $(\mathfrak{p}, \mathfrak{m})=(1, 0)$ exists at
$J = 0$ and $\theta = 0.0887$. This stable fixed point is
surrounded by the stability island in which the quantum
accelerator mode will be created if the atomic initial conditions
are inside the island. The momentum width of the condensate,
$\Delta J_{{\rm BEC}}$ and the stability island $\Delta J_{{\rm
island}}$, are shown with the solid white and black lines
respectively.} \label{unitcell}
\end{figure}
Figure \ref{unitcell} shows a typical phase space portrait for the
map of Eq.\thinspace (\ref{map}) with $\phi_d = 1.4$, $T=29.5
\mu$s and $\epsilon = -0.72$. Perhaps the most important feature
of this plot is the existence of a stable fixed point surrounded
by an island of stability. If the size of these islands is large
enough to capture a significant fraction of the wavepacket they
give rise to observable accelerator modes. According to this model
the momentum gain of an atom in a period $\mathfrak{p}$
accelerator mode after $n$ kicks is given by,
\begin{equation}
q = n \Big [{\eta \tau \over \epsilon} + {2 \pi {\mathfrak{m}}
\over {\mathfrak{p}} | \epsilon |} \Big],
\end{equation}
where ${\mathfrak{m}}$ is an integer and $\mathfrak{(p,m)}$
specifies a particular accelerator mode \cite{Fishman,
Buchleitner, Schlunk}.

The initial conditions of atoms that can be accelerated is another
important property of the accelerator modes. The limited range of
these conditions is a consequence of the fact that the stability
islands do not cover the whole unit cell of the phase space.
Furthermore, the initial momentum required for an accelerator mode
to emerge is periodic. This can be seen by using Eq.\thinspace
(\ref{dynamic}) and the fact that $J$ has a periodicity of $2
\pi$. This is equivalent to a momentum periodicity of $\Delta p =
2 \pi \hbar G /\tau$. Observing this phase space structure
requires that the atomic momentum distribution be narrower than $2
\pi \hbar G /\tau$. This implies a temperature of $\approx 450$ nK
for Rb87 atoms exposed to a pulsed standing wave of 390 nm
wavelength and pulse period close to the Talbot time ($\ell = 2$).
To this extent, a BEC is an ideal candidate \cite{Deng}. The
momentum width of a condensate is shown with two white lines in
Fig.\thinspace \ref{unitcell}. Note that this width is smaller
than the momentum extent of the island, indicating that the
momentum resolution of the experiment is more than sufficient to
clearly detect and identify a stability island from the chaotic
background. However, for experiments utilizing cold thermal atomic
samples, the momentum distribution is significantly wider than
$\hbar G$. Although this wide momentum distribution makes it
relatively easy to observe the accelerator modes, there is no
direct way of examining the structure of the phase space.

In order to use BEC to explore the phase space of the kicked
accelerator we used the experimental setup described in detail by
Ref.\thinspace \cite{ahmadi_2}. A standard six-beam MOT was used
to trap about $50 \times 10^6$ atoms which were loaded into the
trapping potential created by a focused $\mathrm{CO_2}$ laser
beam. In order to optimize the loading efficiency, the waist of
the beam was chosen to be 100 $\mu$m \cite{ahmadi}. Typically
$4\times 10^6$ atoms were trapped at the focus of the
$\mathrm{CO_2}$ laser beam. Subsequently, one of the lenses
through which the beam passed was moved 17 mm in 1 s so as to
reduce the beam waist to 12 $\mu$m and compress the trap for
efficient evaporative cooling \cite{Weiss}. The power in the beam
was then reduced to 100 mW in 4 s to create a pure condensate with
$\sim 50000$ atoms in the $5S_{1/2}F=1,m_F=0$ state. The
$\mathrm{CO_2}$ laser was then turned off and after a variable
time interval the kicking potential was turned on. Varying this
time allowed the BEC to fall under the influence of gravity, thus
changing the momentum of the condensate at the commencement of the
kicks. The light for the kicks was obtained by passing the light
from the MOT laser through a 40 MHz acousto-optic modulator (AOM).
This light was 6.7 GHz red detuned from the $5S_{1/2}F=1
\rightarrow 5P_{3/2}F=2$ transition of Rb87 and propagated at
$41^{\circ}$ relative to the vertical direction. The beam size was
1 mm, such that $\phi_d$ did not change appreciably while the BEC
was interacting with the series of kicks. The phase modulation
depth, $\phi_d$, was inferred by comparing the population of the
first and zeroth order momentum states after one pulse. The
temporal profile of the standing light wave was controlled by
periodically switching the AOM on in order to create a sequence of
pulses each $250$ns in length. The momentum distribution was
measured using a time of flight method. That is, the condensate
expanded for a controlled time interval, typically 9 ms, and was
then destructively imaged using an absorptive technique.
\begin{figure}[h]
\begin{center} \mbox{\epsfxsize 3.0in\epsfbox{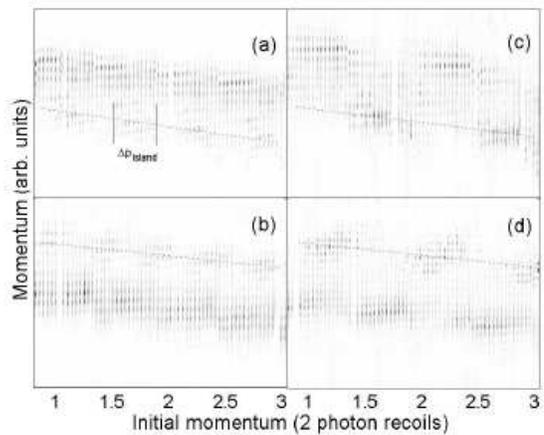}} \end{center}
\caption {Experimental momentum distributions showing the
sensitivity of the QAMs to the initial conditions. The data is
taken after applying 10 kicks with pulse intervals of (a) 61
$\mu$s and (b) 72.2 $\mu$s and 30 kicks with pulse intervals of
(c) 28.5 $\mu$s and (d) 37.1 $\mu$s. The larger number of kicks at
the half-Talbot time was necessary due to the smaller momentum
transfer per pulse. The initial momentum was changed by applying
the kicking potential with variable time delays after releasing
the condensate from the $\mathrm{CO_2}$ laser. The slope of the
data seen in this figure is caused by the momentum gain due to
gravity. The doted lines denote the position of the QAM. Note that
$\Delta p_{{\rm island}}$ is related to the size of the stability
islands by $\Delta p_{{\rm island}}=\hbar G \Delta J_{{\rm
island}} /\tau$. } \label{initial1}
\end{figure}

\begin{figure}
\begin{center} \mbox{\epsfxsize 3.0in\epsfbox{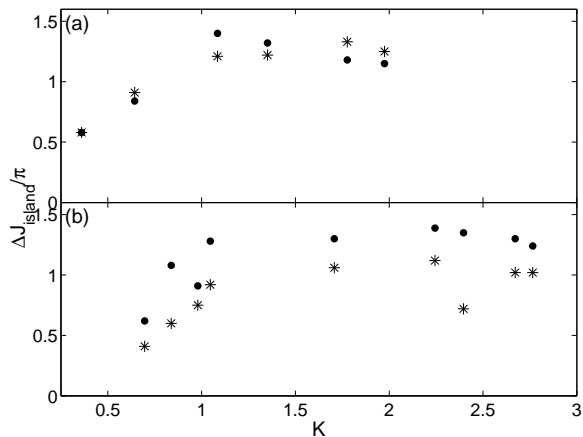}} \end{center}
\caption {Experimental data showing the momentum range in which
the QAMs appear and the corresponding theoretical predictions. (a)
shows the experimental (circle) and theoretical (asterisk) values
for $\Delta J_{{\rm island}}$ near half-Talbot time and (b) show
the same quantities near Talbot time. Figure \ref{initial1}
illustrates the method used to experimentally infer $\Delta
J_{{\rm island}}$.} \label{islandsize}
\end{figure}

To observe the pseudo-classical phase space structure of the QAMs,
a series of data were taken for pulse periods near both Talbot and
half-Talbot times. Figure\thinspace \ref{initial1} shows a typical
data set taken at (a) $T=61 \mu$s, (b) $T=72.2 \mu$s, (c) $T=28.5
\mu$s, and (d) $T=37.1 \mu$s pulse separations for different
values of the BEC's initial momentum. (a) and (b) occur on either
side of the Talbot time at $T = 66.6 \mu$s while (c) and (d) occur
on either side of the half-Talbot time at $T = 33.3 \mu$s. The
data in Figure\thinspace \ref{initial1} was created by
horizontally stacking 60 time-of-flight images of the condensate,
each for a different initial momentum. These data confirm the
periodicity of the QAMs with momentum. Furthermore, the data of
Fig.\thinspace \ref{initial1} provides a direct way to validate
the theoretical prediction of the island size. To do so, the data
of Fig.\thinspace \ref{initial1} was summed along the initial
momentum axis. $\Delta p_{{\rm island}}$ was then determined by
measuring when the accumulated signal of the QAM had dropped to
1/$e$ of its maximum value. The theoretical values were inferred
by plotting the map of Eq.\thinspace (\ref{map}) for the
corresponding experimental values of $K$. The experimental and
theoretical values for the momentum extent of the islands are
given in Fig.\thinspace \ref{islandsize}, near (a) half-Talbot and
(b) Talbot times. The circle and asterisk signs are the
experimental and theoretical values for $\Delta J_{{\rm island}}$
(as defined in Figs.\thinspace \ref{unitcell} and \ref{initial1}).
It can be seen that the experimental values are very close to the
theoretical predictions. Note that for values of $K > 2$ at
half-Talbot time, the stability island elongates in $J$ and
becomes narrow in $\theta$. This behavior reduces the effective
overlap between the BEC's wavefunction and the stability island
and consequently the QAMs were not visible in Fig.\thinspace
\ref{islandsize}(b) for higher values of $K$.
Figure\thinspace \ref{initial1} also shows that there can be little
overlap between the initial conditions that will populate a QAM at
two different values of $T$. This behavior particularly affects what
happens in experiments in which the momentum distribution is
measured as the pulse period is scanned across a resonance time.
Unlike the experiments with cold atomic samples where the QAMs on
both sides of a resonance could be populated \cite{Oberthaler}, in
the case of the condensate, only the QAMs which have significant
overlap with the condensate wavefunction will be observable.
\begin{figure}[h]
\begin{center} \mbox{\epsfxsize 3.0in\epsfbox{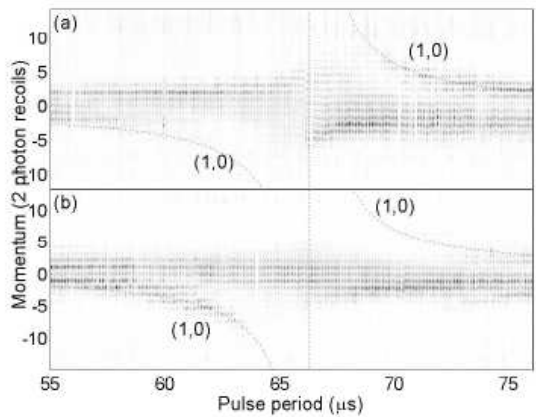}} \end{center}
\caption {Experimental momentum distributions showing controllable
loading of the $\mathfrak{(p,m)}=$(1,0) QAM. The data was taken by
applying 10 kicks for a range of kicking pulse periods. In (a) the
initial momentum was set to $1.2 \hbar G$ so as to efficiently
create the QAM at pulse periods greater than the Talbot time while
in (b) it was set to $1.5 \hbar G$ to efficiently create the QAM
at pulse periods smaller than the Talbot time. The curves show the
position of the QAM calculated with Eq.\thinspace (3).}
\label{period_scane}
\end{figure}
This can be seen in Fig.\thinspace \ref{period_scane}, where we
performed a scan of pulse period across the Talbot time for two
different initial momenta. The initial momentum for Fig.\thinspace
\ref{period_scane}(a) was set to $1.2 \hbar G$ such that the QAMs
were efficiently loaded at pulse periods near $T=72 \mu$s, whereas
in Fig.\thinspace \ref{period_scane}(b), the initial momentum was
set at $1.5 \hbar G$ to mainly populate the QAMs around pulse
periods near $T=61 \mu$s. As can be seen, the QAM with indices
$\mathfrak{(p,m)}=$(1,0) appear at pulse periods greater than the
Talbot time in Fig.\thinspace \ref{period_scane}(a), whereas in
Fig.\thinspace \ref{period_scane}(b) the (1,0) QAM mostly appears
at pulse periods smaller than the Talbot time. Note that this is
the first time that it has been possible to selectively populate
an island at a particular position in phase space.

In conclusion our experiments have demonstrated the feasibility of
observing quantum accelerator modes using a BEC. Using a BEC we were
able to examine the underlying pseudo-classical phase space
structure of the quantum delta kicked accelerator. These experiments
pave the way towards further investigation of more complex systems
like the kicked harmonic oscillator \cite{Gardiner,Duffy} and
dynamical tunneling \cite{Hensinger,Steck} that have received little
study. This experiment also opens the door to experimental
observation of many phenomena related to quantum chaos. For example,
the high momentum resolution of the experiment could bring the
possibility of observing bifurcation of the stability islands to a
practical level. This should lead to a better understanding of the
transition to chaos in a quantum system.

\end{document}